\documentclass[3p,times,procedia]{elsarticle}
\usepackage{nupha_ecrc}

\volume{00}

\firstpage{1}

\journalname{Nuclear Physics A}

\runauth{}

\jid{nupha}

\jnltitlelogo{Nuclear Physics A}

\usepackage{amssymb}

\usepackage[figuresright]{rotating}

\newcommand{\bea}{\begin{eqnarray}}
\newcommand{\eea}{\end{eqnarray}}

\begin{document}

\begin{frontmatter}



\dochead{}

\title{Lattice QCD: bulk and transport properties of QCD matter}


\author[1]{Claudia Ratti}

\address[1]{Physics Department, University of Houston, Houston TX 77204, USA}

\begin{abstract}
We present an overview of the most recent results on bulk and transport properties of QCD matter inferred from lattice QCD simulations.
\end{abstract}


\end{frontmatter}


\section{Introduction}
Lattice QCD is the most reliable first principle tool to address QCD in its non-perturbative regime. Given enough computer power, both statistical and systematic uncertainties can be kept under control. Due to a steady and continuous improvement in computer resources, numerical algorithms and our physical understanding which manifests itself in physical techniques (e.g. the Wilson-flow scale setting introduced in Ref. \cite{Borsanyi:2012zs}), the lattice results which are being produced today reach an unprecedented level of accuracy. This allows a quantitative comparison to experimental observables for the first time in heavy ion physics.
At low temperatures, strongly interacting matter can be well described by a non-interacting gas of hadrons and resonances; in the infinite temperature limit, the system behaves like an ideal, massless gas of quarks and gluons. As we reduce the temperature, interactions between quarks and gluons become relevant: perturbation theory can be systematically used to calculate thermodynamic observables. Resummation techniques improve the convergence of the perturbative series and bring the agreement with lattice QCD results down to $\sim 2.5 T_c$. The temperature range between these two opposite regimes is the realm of lattice QCD: non-perturbative methods are needed to address the relevant observables. This is also the range of temperatures which can be reached in heavy-ion collision experiments: a new synergy between fundamental theory and experiment is today possible due to the precision reached in both approaches. Here we will review the most recent results in the field.
\section{Bulk properties of QCD matter}
\begin{figure}[h!]
\centering
\begin{minipage}{0.45\textwidth}
\includegraphics[width=1.0\textwidth]{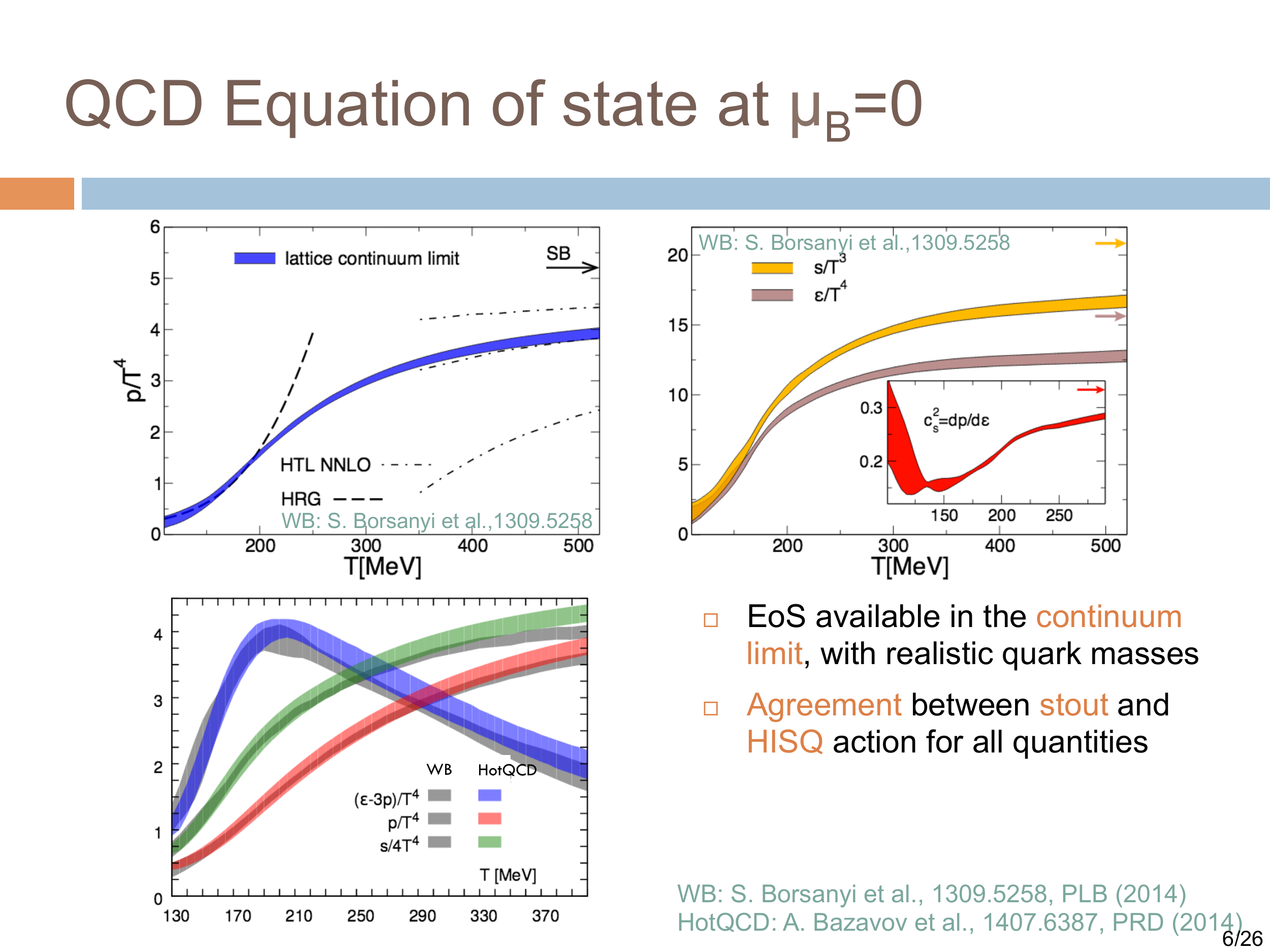}
\end{minipage}
\begin{minipage}{0.45\textwidth}
\includegraphics[width=1.0\textwidth]{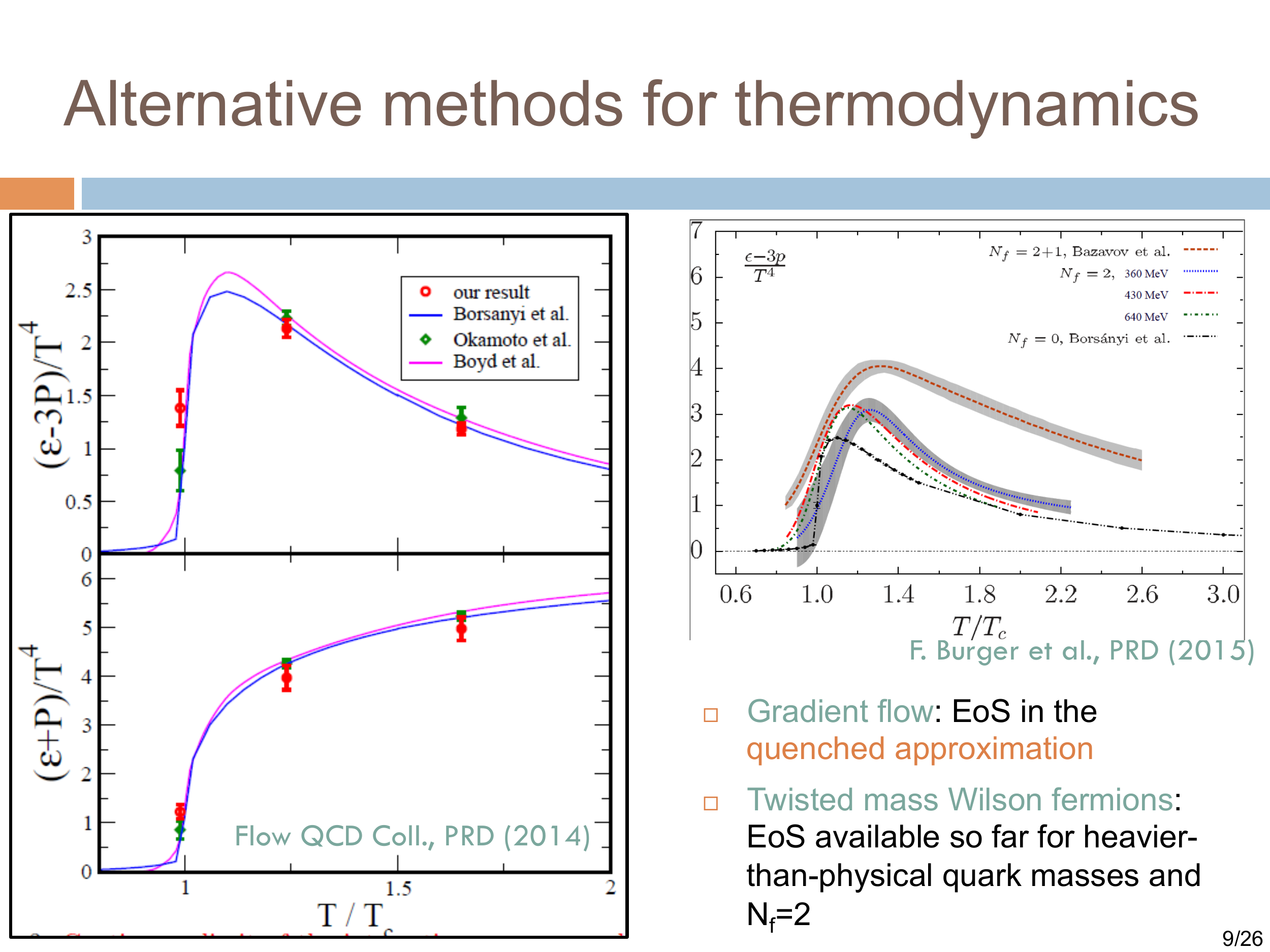}
\end{minipage}
\caption{\label{fig1} Left: comparison between the lattice results of the WB \cite{Borsanyi:2010cj,Borsanyi:2013bia} and HotQCD \cite{Bazavov:2014pvz} collaborations for the equation of state of QCD. Right: Comparison between the equations of state of pure gauge QCD from Refs. \cite{Burger:2014xga,Borsanyi:2012ve} and the one of 2+1 flavor QCD \cite{Bazavov:2014pvz}.}
\end{figure}

The most reliable results obtained from lattice QCD simulations concern thermodynamic observables in equilibrium. For example, the equation of state of QCD is now available for a system of 2+1 dynamical quark flavors with physical quark masses in the continuum limit. In 2014, the HotQCD collaboration published continuum results for pressure, energy density, entropy density and interaction measure as functions of the temperature \cite{Bazavov:2014pvz} which agree with the ones previously obtained by the Wuppertal Budapest (WB) collaboration \cite{Borsanyi:2010cj,Borsanyi:2013bia}: they are showed in the left panel of Fig. \ref{fig1}. These results have been independently obtained with two different staggered fermion actions (2stout and HISQ): the agreement between the two is a fundamental test of the validity of the discretized lattice method to solve QCD. Recently, first results for the equation of state obtained from other approaches to lattice QCD are becoming available: these include the gradient flow method \cite{Asakawa:2013laa}, which extracts the thermodynamic quantities from the energy-momentum tensor, and twisted mass fermions \cite{Burger:2014xga}; the former are limited so far to the quenched approximation, the latter to two flavors with heavier-than-physical quark masses.

The lattice results for QCD thermodynamics are so far limited to zero or small chemical potentials, due to the ``sign problem", which makes direct simulations of QCD at finite density not feasible. Some promising alternative methods are becoming available, but their application to QCD with
physical parameters and controlled discretization has not yet been achieved \cite{Aarts:2013uxa,Cristoforetti:2013wha}. Other methods have been proposed to circumvent the sign problem: here we will focus on the Taylor expansion of thermodynamic observables around $\mu_B=0$ \cite{Allton:2002zi,Gavai:2008zr} (which can be considered as a truncated version of the multiparameter reweighting \cite{Fodor:2001au}) and analytic continuation from imaginary chemical potentials \cite{deForcrand:2002hgr,D'Elia:2002gd,Wu:2006su}. 

\begin{figure}[h!]
\centering
\includegraphics[width=1.0\textwidth]{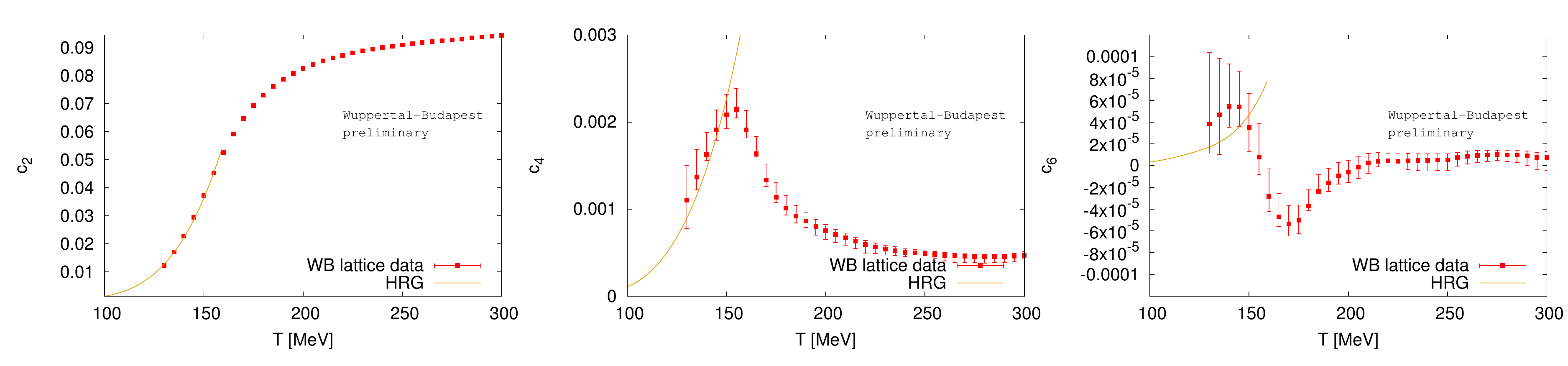}
\caption{\label{fig2} Preliminary results for the Taylor coefficients $c_2...c_6$ as functions of the temperature from the WB collaboration, obtained from imaginary $\mu_B$ simulations. The data are continuum extrapolated; the error-bars are only statistical: the systematics of the $\mu_B$ fitting are not included \cite{Sz}.}
\end{figure}

One can expand the QCD pressure in Taylor series around $\mu_B=0$:
\bea
\frac{p(\mu_B)}{T^4}=c_0(T)+c_2(T)\left(\frac{\mu_B}{T}\right)^2+
c_4(T)\left(\frac{\mu_B}{T}\right)^4+c_6(T)\left(\frac{\mu_B}{T}\right)^6+\mathcal{O}(\mu_B^8);
\label{Taylor}
\eea
this expansion contains the coefficients $c_i(T)$, extracted from lattice QCD simulations. After the early results for $c_2...c_6$ \cite{Allton:2005gk}, the first continuum extrapolated results for $c_2$ were published in Ref. \cite{Borsanyi:2012cr}; in Ref. \cite{Hegde:2014sta} $c_4$ was showed, but only at finite lattice spacing. In Fig. \ref{fig2} we show the preliminary, continuum extrapolated results for $c_2,~c_4$ and $c_6$ as functions of the temperature. Such results have been obtained by the WB collaboration from imaginary $\mu_B$ simulations: $c_2,...c_6$ have been fitted on the $\mu_B-$derivatives of $p/T^4$ for fixed temperature \cite{Sz}.

\begin{figure}[h!]
\centering
\begin{minipage}{0.45\textwidth}
\includegraphics[width=1.0\textwidth]{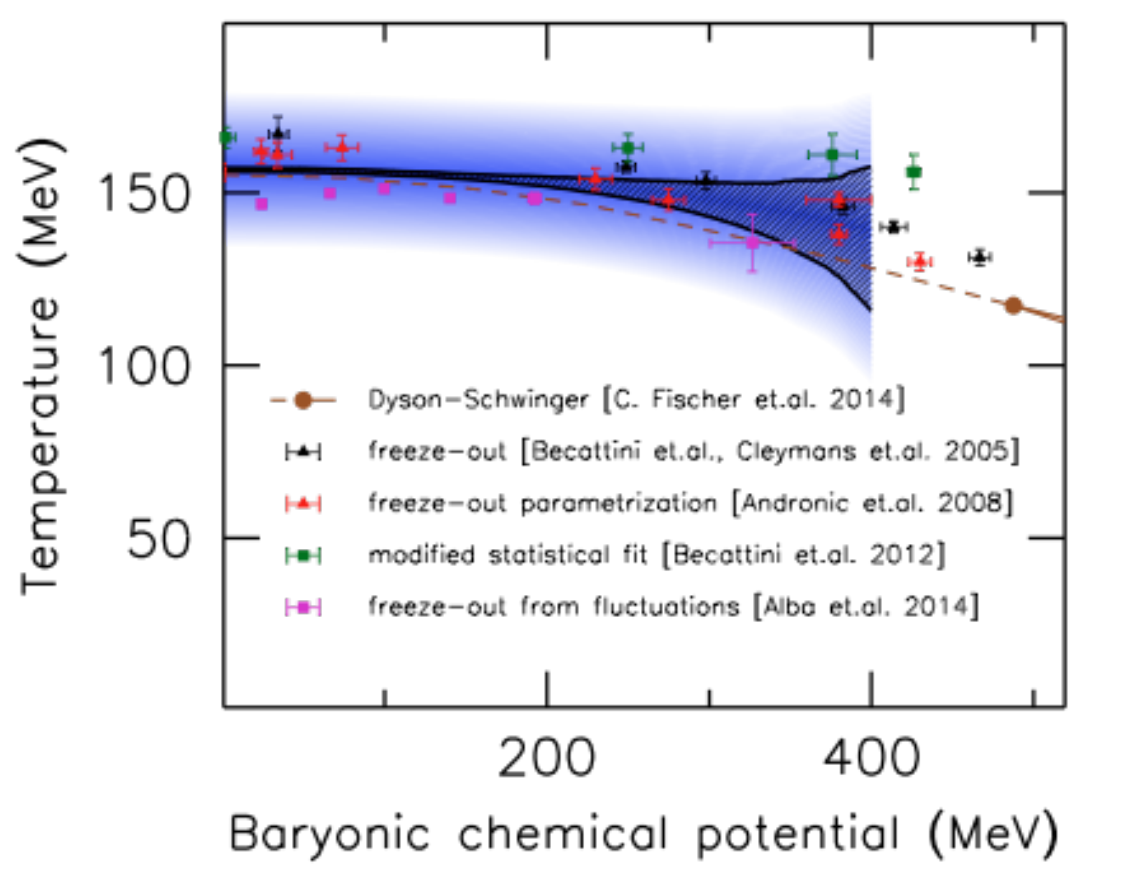}
\end{minipage}
\begin{minipage}{0.45\textwidth}
\includegraphics[width=1.0\textwidth]{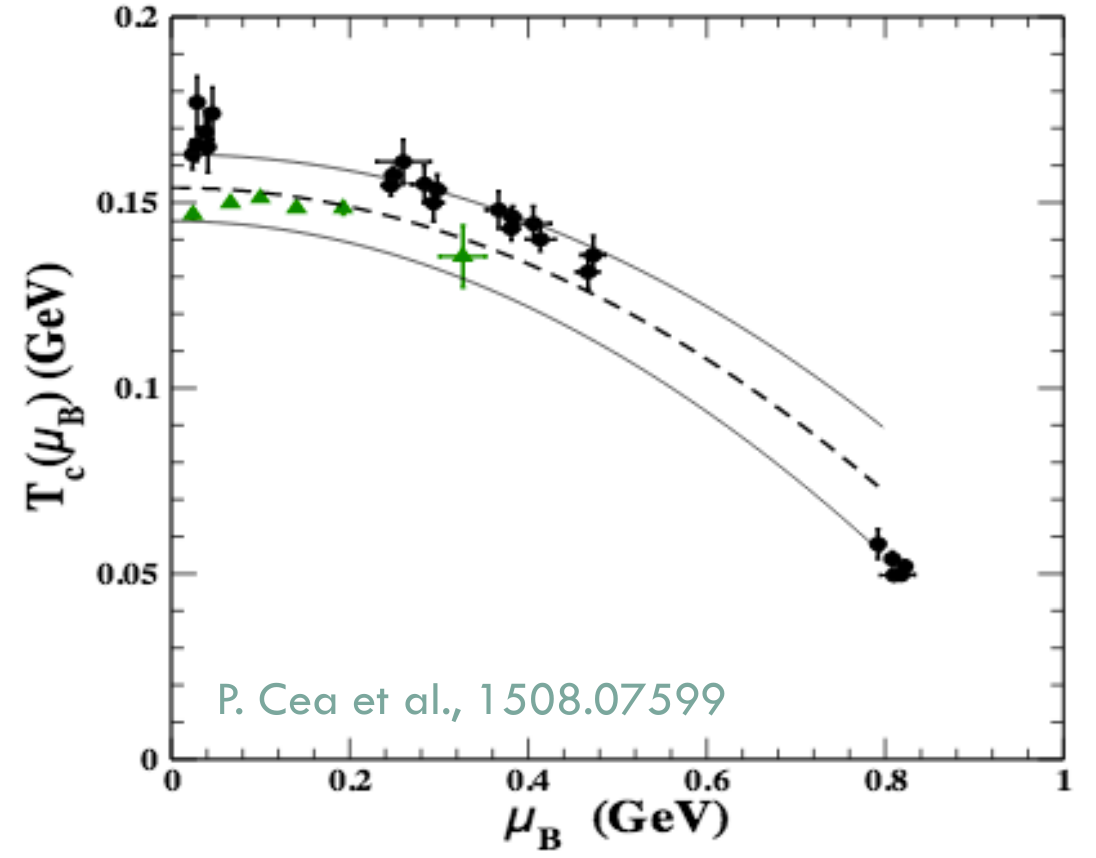}
\end{minipage}
\caption{\label{fig3} Left: The  phase  diagram  based  on  the $\mu_B-$dependent $T_c$ from  the  chiral  condensate, analytically continued from imaginary chemical potential \cite{Bellwied:2015rza}.  The blue band indicates the width of the transition. 
The shaded black region shows the transition line obtained from the chiral
condensate.  The widening around 300 MeV is coming from the uncertainty of the curvature
and  from  the  contribution  of  higher  order  terms,  thus  the  application  range  of  the  results
is  restricted  to  smaller values. We also  show  some
selected non-lattice results:  the Dyson-Schwinger result \cite{Fischer:2014ata}, and the freeze-out data of Refs. \cite{Cleymans:2004pp}-\cite{Alba:2014eba}. Right: analogous plot from Ref. \cite{Cea:2015cya}.}
\end{figure}

At $\mu_B=0$, the QCD phase transition is an analytic crossover \cite{Aoki:2006we}; a pseudocritical temperature $T_c$ can be defined by looking at the inflection point or peak of some specific observables \cite{Aoki:2006br,Aoki:2009sc,Borsanyi:2010bp,Bazavov:2011nk}. One can follow the change in their position as the chemical potential increases: this gives rise to a $\mu_B-$dependence of $T_c$ which can be expressed as:
\bea
\frac{T_c(\mu_B)}{T_c(\mu_B=0)}=1-\kappa\left(\frac{\mu_B}{T_c(\mu_B)}\right)^2+\lambda\left(\frac{\mu_B}{T_c(\mu_B)}\right)^4+...~.
\label{curvature}
\eea
The parameter $\kappa$ in the above expansion is the curvature of the phase diagram and it can be extracted from lattice QCD simulations: by looking at three different observables (chiral condensate, chiral susceptibility and strange quark susceptibility) the WB collaboration recently published a value of $\kappa=0.0149\pm0.0021$ \cite{Bellwied:2015rza}; this has been obtained by fixing the strange quark chemical potential to impose strangeness neutrality. The phase diagram corresponding to this value of $\kappa$ is showed in the left panel of Fig. \ref{fig3}, together with a compilation of freeze-out parameters obtained with different methods. Similar results have been obtained recently by two other groups: P. Cea {\it et al.} obtain a value of $\kappa=0.020(4)$ by fixing $\mu_s=\mu_l$ \cite{Cea:2015cya}, while Bonati {\it et al.} find $\kappa=0.0135(20)$ both with $\mu_s=0$ and $\mu_s=\mu_l$ \cite{Bonati:2015bha}.  In the right panel of Fig. \ref{fig3}, the phase diagram with the curvature from Ref. \cite{Cea:2015cya} is shown.

\begin{figure}[h!]
\centering
\begin{minipage}{0.43\textwidth}
\includegraphics[width=1.2\textwidth]{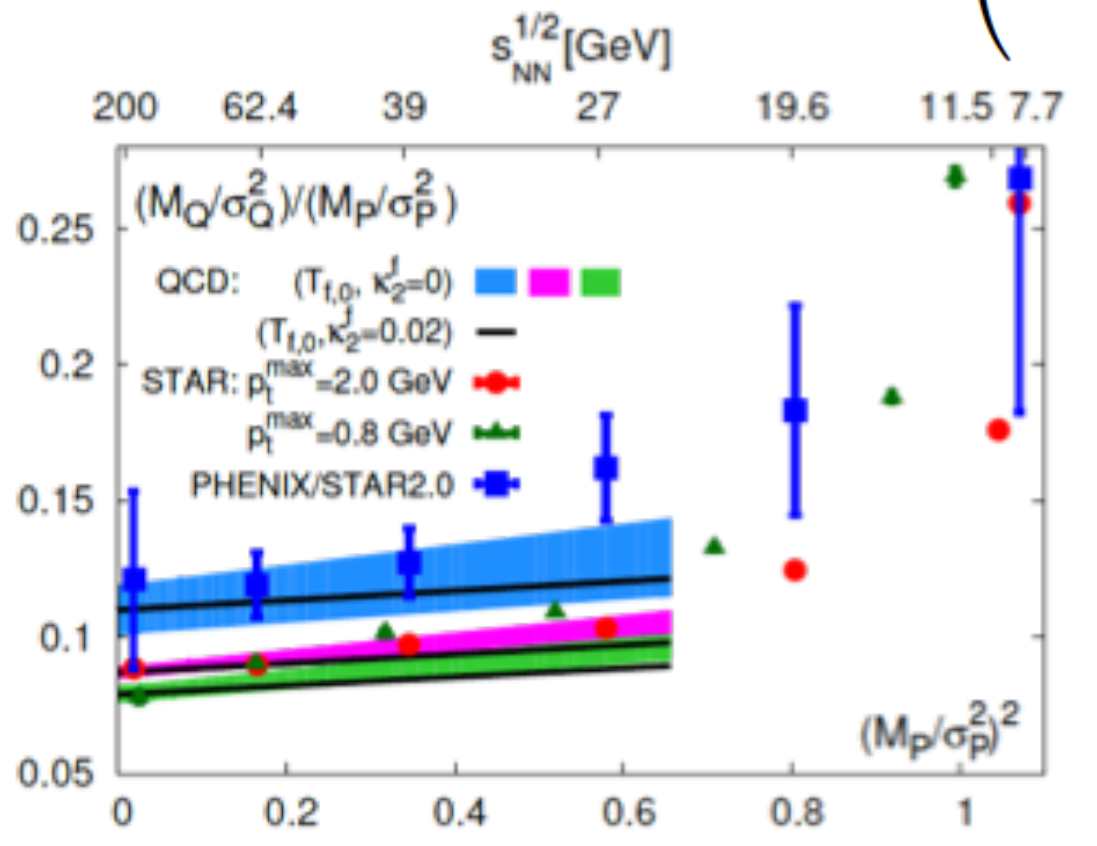}
\end{minipage}
\hspace{1.1cm}
\begin{minipage}{0.45\textwidth}
\includegraphics[width=1.0\textwidth]{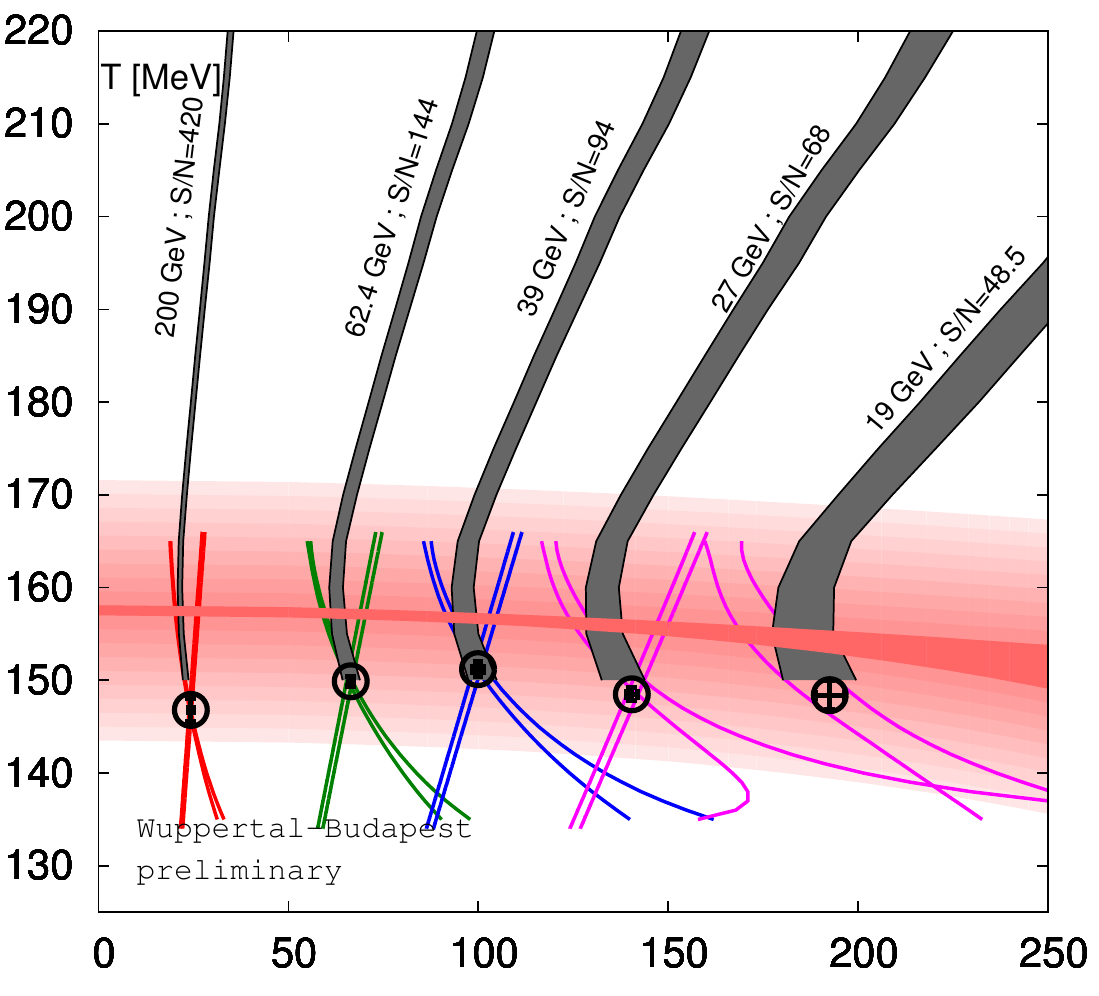}

\end{minipage}
\caption{\label{fig4} Left: From Ref. \cite{Bazavov:2015zja}: the ratio of ratios of $\chi_1/\chi_2$ for net electric charge and net-proton fluctuations measured by the STAR and PHENIX Collaborations \cite{Adamczyk:2013dal,Adamczyk:2014fia,Luo,Adare:2015aqk}. Right: Preliminary results of the WB collaboration. The colored full and dashed lines are the contours at constant mean/variance ratios of the net electric charge from lattice simulations. The contours that correspond to STAR data intersect in the freeze-out points of Ref. \cite{Alba:2014eba}. The red band is the QCD phase diagram shown in Fig. \ref{fig3}. Also shown are the isentropic contours that match the chemical freeze-out data \cite{Sz}.}
\end{figure}

Among the most interesting observables which can be simulated on the lattice are fluctuations of conserved charges; they are defined as derivatives of the pressure with respect to the chemical potentials of conserved charges (baryon number $B$, electric charge $Q$, strangeness $S$):
\bea
\chi_{lmn}^{BQS}(T,\mu_B)=\frac{\partial^{l+m+n}p/T^4}{\partial(\mu_B/T)^l\partial(\mu_Q/T)^m\partial(\mu_S/T)^n}.
\label{fluctuations}
\eea
The $\mu_B=0$ diagonal second-, fourth- and sixth-order baryon number fluctuations are the Taylor expansion coefficients of Eq. (\ref{Taylor}), shown in Fig. \ref{fig1}. Their interest resides in the fact that the lattice results can be compared to experimental measurements, to the purpose of extracting information on the QCD matter created in heavy-ion collisions: while higher order fluctuations can be used to gain information about the position of the critical point in the QCD phase diagram \cite{Stephanov:1999zu,Gavai:2008zr}, the lower order ones can lead to the determination of the freeze-out temperature and chemical potential in the evolution of the system, at which all inelastic reactions cease \cite{Karsch:2012wm}-\cite{Borsanyi:2013hza}. Indeed, the fluctuations of a given conserved charge are the cumulants of its event-by-event distribution; volume-independent ratios can conveniently be defined, which allow to determine the freeze-out temperature and chemical potential by comparing the lattice QCD curves to the experimental value. For a meaningful comparison, all non-thermal sources of fluctuations must be understood and kept under control, and a variety of effects has been identified and studied in the literature \cite{Skokov:2012ds}-\cite{Becattini:2012xb}. In 2014 the WB collaboration found that, analyzing the fluctuations of electric charge and baryon number independently, there is a consistency between the freeze-out chemical potentials corresponding to the highest RHIC energies \cite{Borsanyi:2014ewa,Adamczyk:2013dal,Adamczyk:2014fia}. Recently, the authors of Ref. \cite{Bazavov:2015zja} performed a fit to the ratio of ratios of $\chi_1/\chi_2$ (mean/variance) for electric charge an proton number and were able to obtain both the freeze-out temperature and the curvature of the freeze-out line. The value of the freeze-out temperature ($T_f=(147\pm2)$ MeV) is in agreement with the one obtained in Ref. \cite{Borsanyi:2014ewa}. The left panel of Fig. \ref{fig4} shows the ratio of ratios of $\chi_1/\chi_2$ for electric charge and proton number used for this fit. Along the same lines, the WB collaboration performed a combined fit of $\chi_1/\chi_2$ for electric charge and proton number and found the freeze-out temperature and chemical potential for the highest RHIC energies. These preliminary results are shown in the right panel of Fig. \ref{fig4}, together with the isentropic lines which match the freeze-out data, the contours for constant mean/variance of net-electric charge from the lattice, and the results of a previous analysis based on the HRG model \cite{Alba:2014eba}. The WB results agree with the ones of Ref. \cite{Bazavov:2015zja} and with the HRG model ones. Unfortunately, fluctuation data are not yet available at the LHC. However the authors of Ref. \cite{Braun-Munzinger:2014lba}, assuming that the lower moments follow a Skellam distribution, expressed the second moments in terms of the particle yields and compared the lattice results to the ALICE experimental data, finding a slightly higher freeze-out temperature than then ones obtained at RHIC. Recently, the study of fluctuations has been extended to very large temperatures \cite{Bellwied:2015lba,Ding:2015fca} to extract the onset of the HTL perturbative expansion \cite{Haque:2014rua,Mogliacci:2013mca}, which is found to be $T\simeq 250$ MeV. 
Fluctuations are also useful to infer the degrees of freedom which populate the Quark-Gluon Plasma (QGP) around the transition temperature. For example, studying the correlations between charm and baryon number, it is possible to extract the temperature at which the charm quarks are liberated. A recent study \cite{Mukherjee:2015mxc} shows that, even if the onset of deconfinement for the charm quark takes place around $T\simeq165$ MeV, it becomes the dominant degree of freedom in the thermodynamics of the charm sector only at $T\simeq200$ MeV, while between these two temperatures the dominant contribution to the charmed pressure is given by open charm meson- and baryon-like excitations with integral baryonic charge.
\section{Transport properties of QCD matter}
In the region around 1-2 $T_c$, QCD matter is highly non-perturbative and significant modifications of its transport properties are expected. Unfortunately, the observables that are related to the transport properties of matter all have one common feature that makes it difficult to extract them from lattice QCD simulations. The latter can investigate a certain set of current-current correlators on a discrete set of points. Such correlators have a spectral representation which involves integrals of spectral functions weighted by appropriate integration kernels. Extracting the desired observables (the low-frequency and low-momentum limit of such spectral functions) requires the application of inversion methods or a modeling of the spectral functions at low frequencies in order to integrate over a discrete set of lattice points. In spite of these difficulties, several results have been obtained recently on the transport properties of matter.

\begin{figure}[h!]
\centering
\begin{minipage}{0.43\textwidth}
\includegraphics[width=1.2\textwidth]{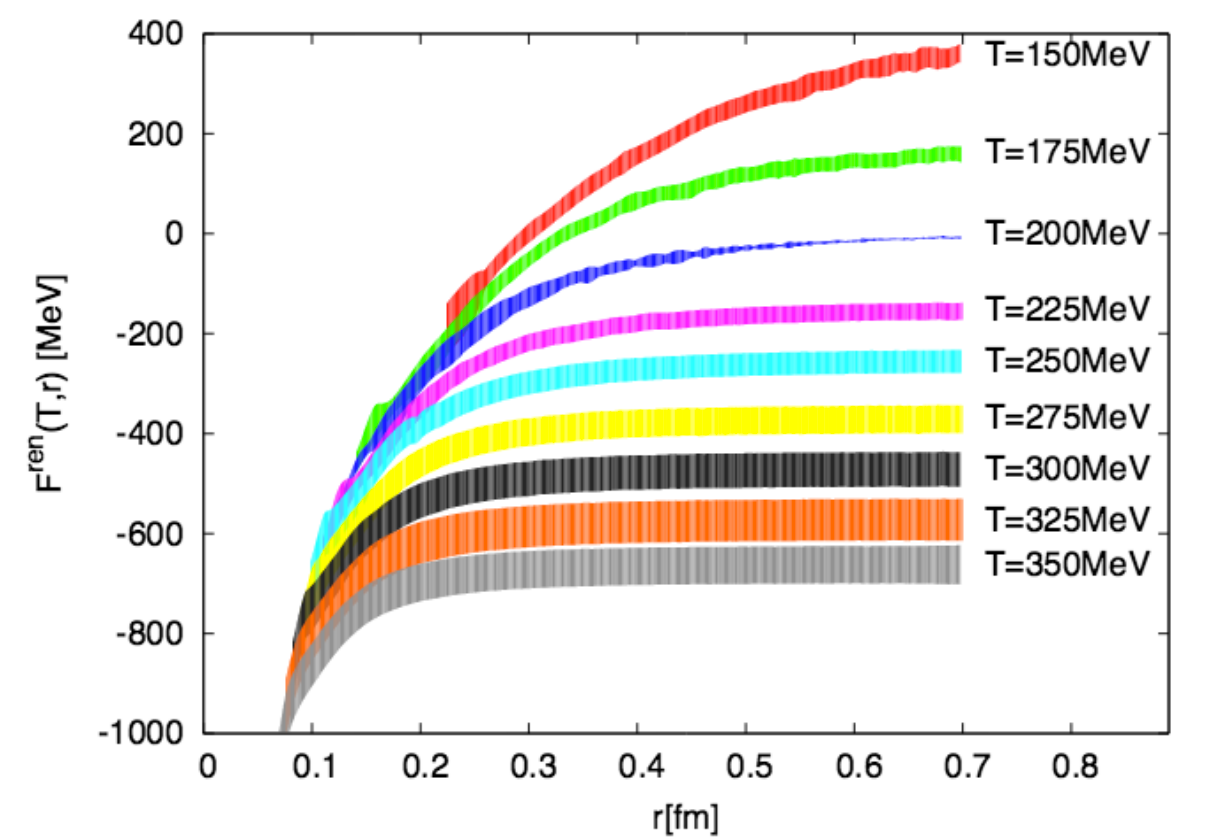}
\end{minipage}
\hspace{1.1cm}
\begin{minipage}{0.45\textwidth}
\includegraphics[width=1.0\textwidth]{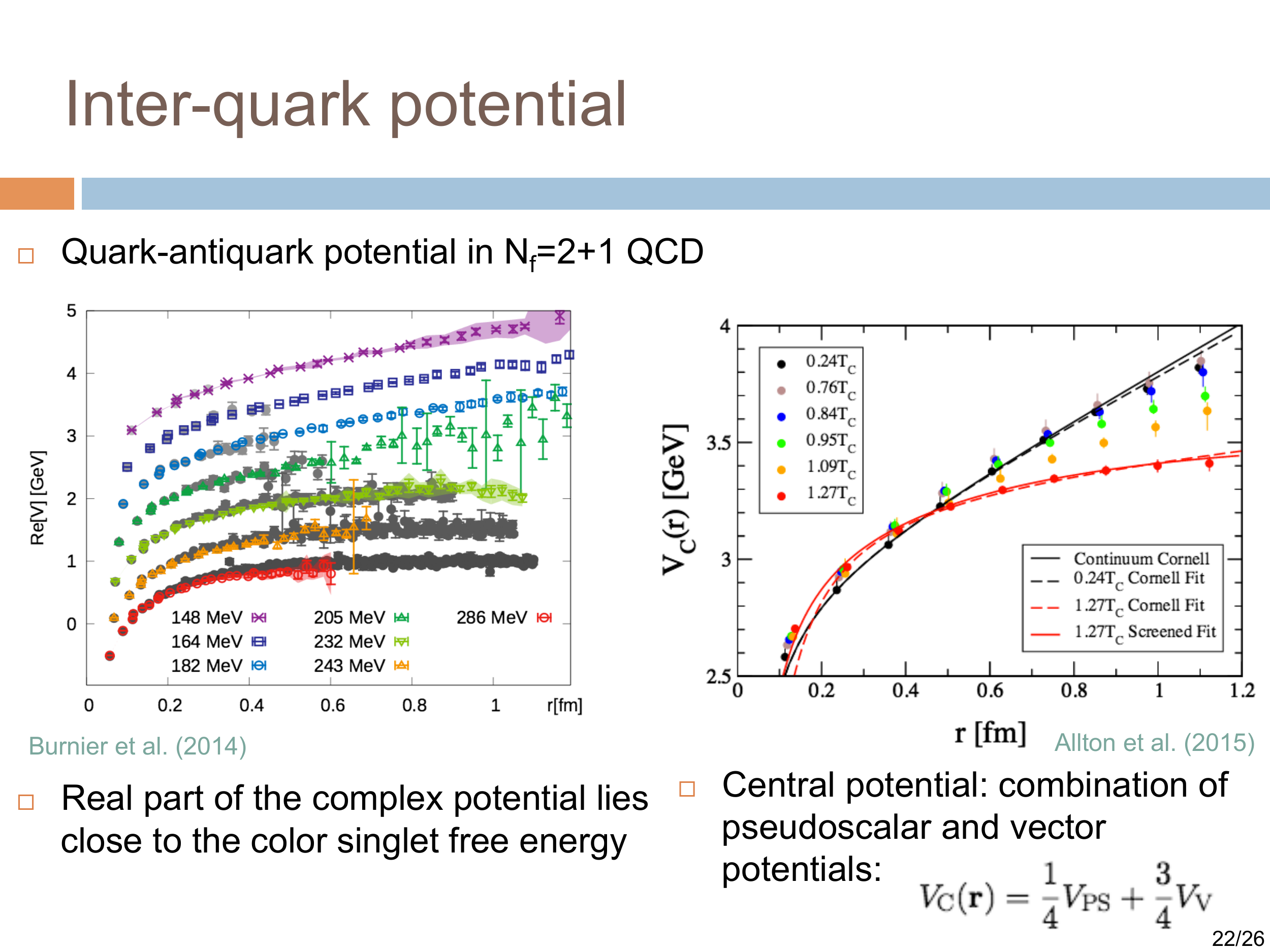}
\end{minipage}
\caption{\label{fig5} Left: From Ref. \cite{Borsanyi:2015yka}: continuum values for the static $q\bar{q}$ free energy at different temperatures for a system of 2+1 quark flavors with physical masses. Right: from Ref. \cite{Burnier:2014ssa}: the real part of the static interquark potential (open symbols) compared to the color singlet free energies in Coulomb gauge (gray circles). }
\end{figure}

One of the most interesting points concerns the properties of quarkonia. Usually, this problem is addressed by means of three distinct approaches:
\begin{itemize}
\item{extract the quark-antiquark potential and plug it into Schr\"odinger's equation for the bound state two-point function}
\item{extract the quarkonia spectral functions from euclidean temporal correlators}
\item{study spatial correlators and their in-medium screening properties.}
\end{itemize}
Here I will concentrate on the first two. For the first approach, we have continuum extrapolated results for the $q\bar{q}$ free energy obtained from correlators of two Polyakov loops: these results have been obtained for a system of 2+1 flavors at the physical mass \cite{Borsanyi:2015yka}: they are shown in the left panel of Fig. \ref{fig5}. Also, the $q\bar{q}$ potential has been obtained in a system of 2+1 dynamical quark flavors using a new Bayesian inference prescription \cite{Burnier:2014ssa}: these results are shown in the right panel of Fig. \ref{fig5}. Other results have been obtained by assuming the validity of Schr\"odinger's equation for charm quarks and extracting the potential directly from charmonium correlators \cite{Allton:2015ora}. Both methods agree with each other and show the typical Debye-screening flattening of the potential at high temperatures.

\begin{figure}[h!]
\centering
\includegraphics[width=0.9\textwidth]{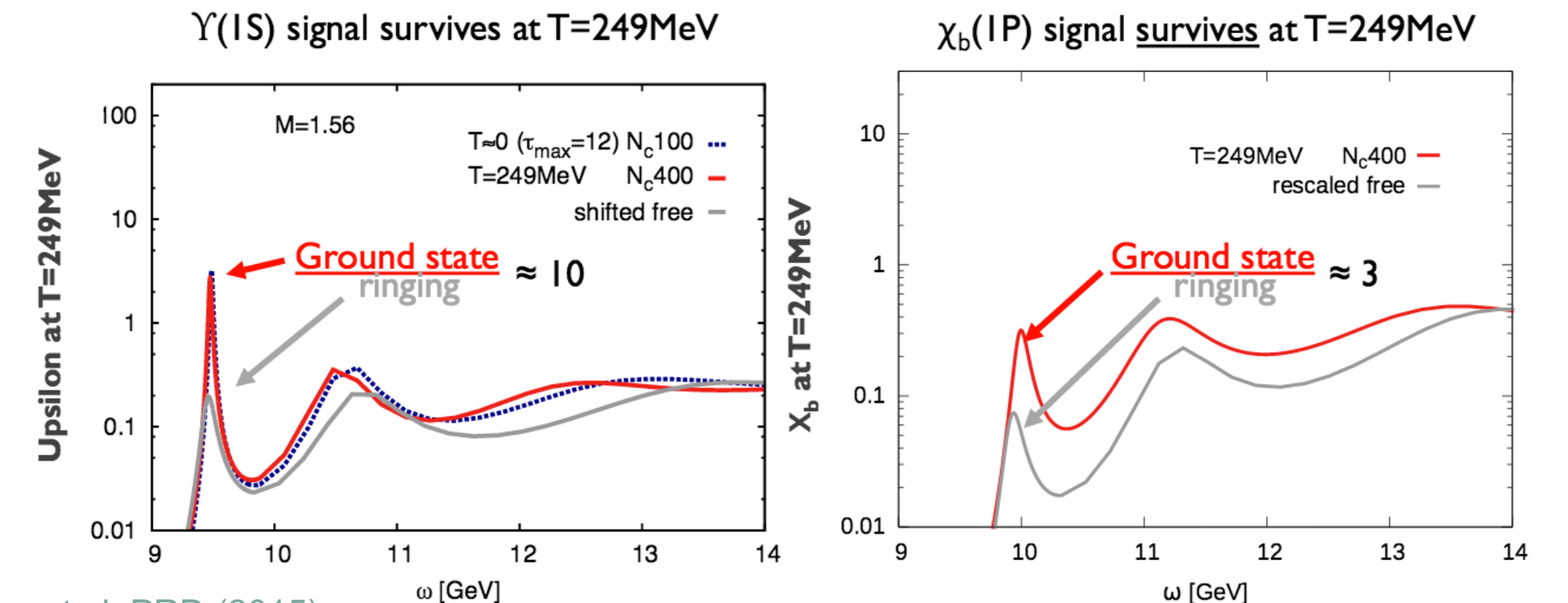}
\caption{\label{fig7} Spectral functions for the s- (left) and p- (right) wave bottomonium states \cite{Rothkopf}.}
\end{figure}

As for the charmonium spectral functions, both the quenched approximation results and the ones with dynamical quarks show that all charmonium states are dissociated at $T\geq 1.5T_c$ \cite{Aarts:2007pk}-\cite{Borsanyi:2014vka}. The situation is different for the bottomonium, for which there is a discrepancy between different analyses. G. Aarts and his collaborators, using the Maximum Entropy Method to reconstruct the spectral function, show that the s-wave state survives up to $T\simeq 1.9T_c$ while the p-wave one melts just above $T_c$ \cite{Aarts:2014cda}. By using the Bayesian method to reconstruct the spectral function, S. Kim {\it et al.} find that both s- and p-waves survive in the plasma up to $T\simeq250$ MeV \cite{Kim:2014iga}: these findings are shown in Fig. \ref{fig7} \cite{Rothkopf}.

\begin{figure}[h!]
\centering
\begin{minipage}{0.4\textwidth}
\includegraphics[width=1.2\textwidth]{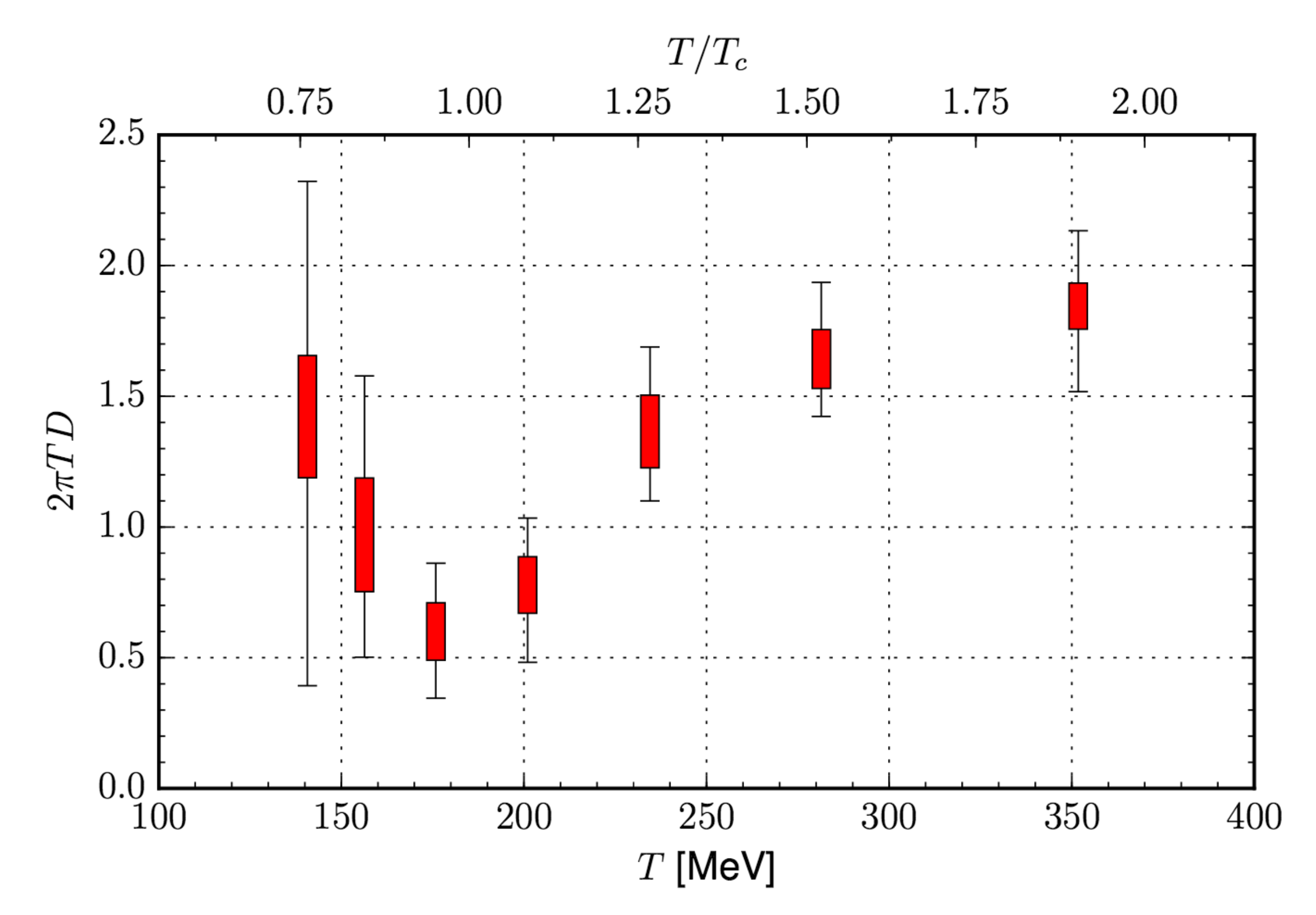}
\end{minipage}
\hspace{1.1cm}
\begin{minipage}{0.5\textwidth}
\includegraphics[width=1.0\textwidth]{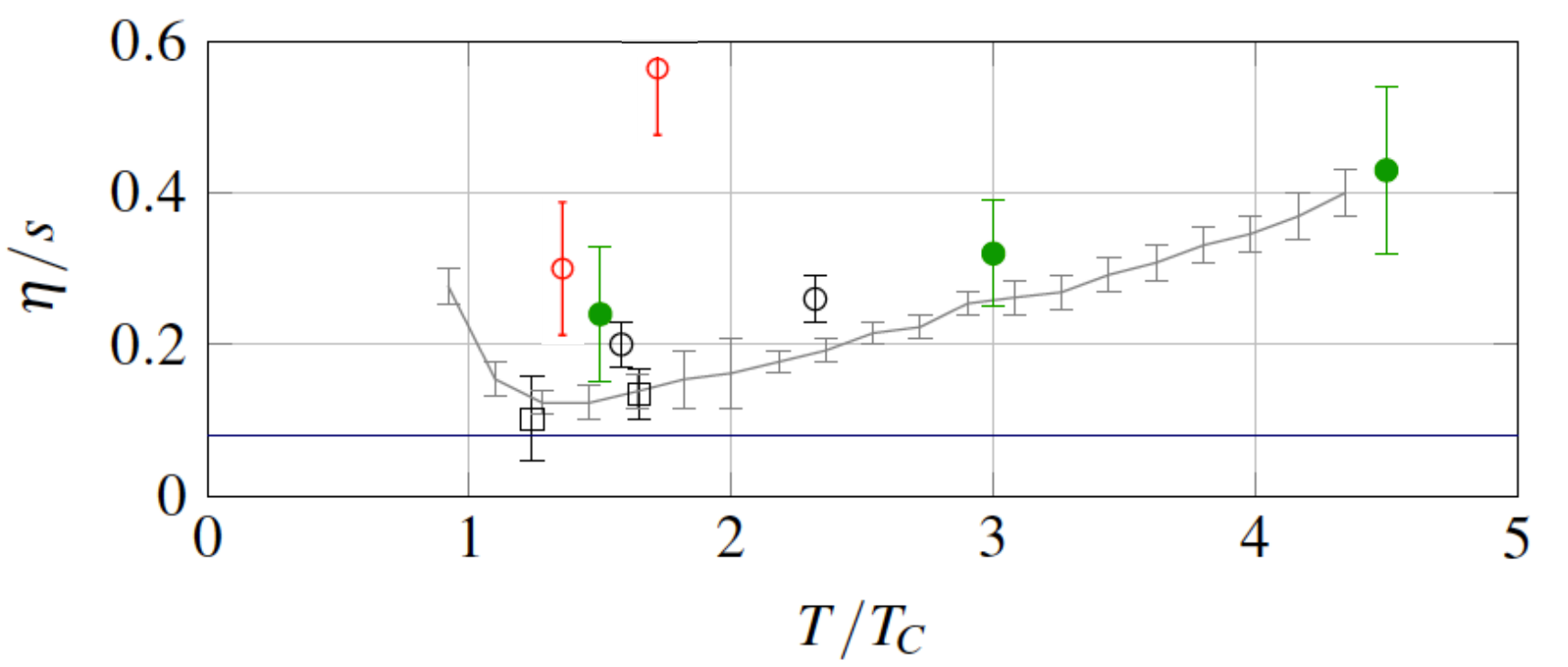}
\end{minipage}
\caption{\label{fig6} Left: From Ref. \cite{Aarts:2014nba}: diffusion coefficient $D$ multiplied by $2\pi T$ as a function of the temperature, using $D=\sigma/\chi_2^Q$. Right: compilation of all available lattice QCD results on the pure gauge shear viscosity over entropy as a function of the temperature: H. Meyer (black squares and circles) \cite{Meyer:2007dy,Meyer:2009jp}, Christiansen {\it et al.} (vertical lines)\cite{Christiansen:2014ypa}, Nakamura and Sakai (empty red circles) \cite{Nakamura:2004sy}, S. W. Mages {\it et al.} (green full circles) \cite{Mages:2015rea}.}
\end{figure}

One of the transport coefficients that have been extensively studied on the lattice is the electric conductivity, $\sigma$, which measures the response of the medium to small perturbations induced by an electromagnetic field. Several results are available for this quantity \cite{Ding:2010ga,Brandt:2012jc}, but so far only one has been obtained in a system of 2+1 quark flavors in Ref. \cite{Aarts:2014nba}, by means of the Maximum Entropy Method. The electric conductivity increases by a factor 6 in the range of temperatures between 140 and 350 MeV. The charge diffusion coefficient has also been obtained, by dividing the electric conductivity by the second order fluctuation $\chi_2^Q$. The diffusion coefficient has a dip close to $T_c$, which is consistent with the expectations of a strongly coupled system. These results are shown in the left panel of Fig. \ref{fig6}.

The right panel of Fig. \ref{fig6} shows a compilation of all available lattice QCD results on the pure gauge shear viscosity over entropy ratio as a function of the temperature. For this observable, besides the difficulty of inverting the energy-momentum tensor correlator, an additional problem arises: this correlator itself is extremely noisy, and no technique is available to reduce it if quarks are introduced in the simulations. This is the reason why so far only quenched results are available. An algorithm which allows to increase the signal-to-noise ratio is needed to extract this observable also in the full QCD case.



\section{Conclusions}
As shown by the large amount of new results summarized in these proceedings, which became available since the 2014 Quark Matter conference, the progress and the precision achieved by lattice QCD simulation is really impressive. Precise results are available for QCD thermodynamics at zero and small chemical potentials, which allow a quantitative comparison with experimental results for the first time. Progress has been made in the determination of real time dynamics. This should enable us to achieve a comprehensive understanding of bulk and transport properties of QCD matter from lattice QCD simulations.
\section*{Acknowledgements}
I would like to thank all my lattice QCD colleagues who sent me their contributions and comments to my plenary talk. This work is supported by the National Science Foundation through grant  number  NSF  PHY-1513864 and the DOE INCITE program supported under Contract DE-AC02-06CH11357.
\bibliographystyle{elsarticle-num}
\bibliography{<your-bib-database>}



\end{document}